\documentclass[aps,showpacs,showkeys,preprint]{revtex4}
\usepackage{amssymb}
\usepackage{graphicx}
\newcommand{\half}{\frac{1}{2}}
\newcommand{\ep}{\epsilon}

\begin{document}
 \preprint{Alberta Thy  11-06}

\title{Large mass expansion in two-loop QCD corrections of paracharmonium decay}



\author{~K.~Hasegawa}
\email[]{hasegawa@phys.ualberta.ca}
\author{~Alexey~Pak}
\email[]{apak@phys.ualberta.ca}
\affiliation{Department of Physics, University of Alberta, Edmonton, Alberta T6G 2J1, Canada}



\date{\today}

\begin{abstract}
We calculate the two-loop QCD corrections to paracharmonium  decays $\eta_{c} \rightarrow
\gamma \gamma$ and $\eta_{c} \rightarrow g g$ involving light-by-light  scattering diagrams 
with light quark loops. Artificial large mass expansion and convergence improvement techniques 
are used to evaluate these corrections. The obtained corrections to the decays $\eta_{c} \rightarrow
\gamma \gamma$ and $\eta_{c} \rightarrow g g$ account for $-1.25 \%$ and  $ -0.73 \%$ of the 
leading order contribution, respectively.
\end{abstract}
 

\pacs{12.38.Bx, 12.38.-t, 13.20.Gd}
\keywords{Large mass expansion, charmonium, QCD}

\maketitle


\section{Introduction}

Since the year 1974, when bound $c\bar{c}$ states (charmonium) were discovered, 
they have been thoroughly studied experimentally and theoretically, contributing to better knowledge of the 
Standard Model parameters. Generally, the theoretical description of charmonium decays requires 
the knowledge of the perturbative effects in $c\bar{c}$ annihilation, and non-perturbative corrections to 
the charmonium wave function. However, ratios of decay rates into different final states are only sensitive 
to perturbative QCD contributions.  
In the present paper we focus on the subset of two-loop perturbative corrections to the ground-state paracharmonium 
decays $\eta_{c}\to\gamma\gamma$ and $\eta_{c}\to g g$. Abelian $\mathcal{O}(\alpha_s)$
corrections to both decays are related to the one-loop correction to parapositronium decay by a trivial 
replacement of coupling constants \cite{har}. Non-abelian one-loop contributions were presented in \cite{bar, hag}. 
$\mathcal{O}(\alpha_s^2)$ corrections to $\eta_{c}\to\gamma\gamma$ were discussed in \cite{cza1}, 
but those results did not include light-by-light scattering type contributions. Then we calculate these diagrams which
form a gauge-independent subset.

Multiloop corrections in the perturbation theory can be calculated by the asymptotic expansion \cite{smi1}.  
But at the present stage, it is impossible to calculate the two-loop diagrams considered in this paper by the ordinary 
asymptotic expansions because of the complexity of the diagrams. In order to calculate them we adopt the 
method of the large mass expansion (LME) \cite{smi2, tka}. We introduce an artificial large mass M for the
internal charm quarks and execute the asymptotic expansion with the mass scale of the external charm 
quarks $m_{c}$ (soft scale) and the mass scale of the internal charm quarks M (hard scale).
When the loop momenta of the internal charm quarks are factorized into the soft  scale, we can expand
the propagators of the charm quarks by the ratio $m_{c}/\mbox{M}$. The expansions of the propagators
reduce the original diagrams to the simpler ones which we can calculate. 
After that we obtain the results in a series of $m_{c}/\mbox{M}$ and identify the artificial mass M with
the original one $m_{c}$.

The present paper is organized as follow. 
In Sec. \ref{sec2},  we review the example of parapositronium decay $\mbox{p-Ps}\to \gamma\gamma$
to demonstrate the method of LME and obtain two-loop QED results.
In Sec. \ref{sec3}, we adapt the results in the p-Ps decay to the paracharmonium decays
$\eta_{c} \rightarrow \gamma \gamma$ and $\eta_{c} \rightarrow g g$, and obtain the two-loop QCD
corrections by counting the color factors. In Sec. \ref{sec4}, we have a summary.

\section{Parapositronium decay   \label{sec2}}
\begin{figure}[tb]
\begin{center}
\includegraphics[width=6cm]{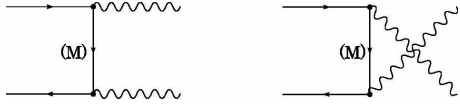}
\caption{Two diagrams which contribute to the p-Ps decay at tree level.    \label{fig1}}
\end{center}
\end{figure}

\begin{figure}[tb]
\begin{center}
\includegraphics[width=7cm]{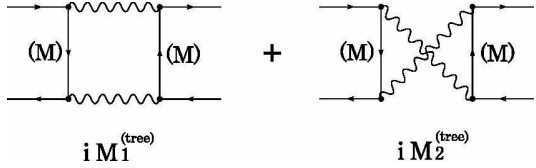}
\end{center} 
\caption{Two diagrams $\mbox{M}_{1}^{tree}$ and $\mbox{M}_{2}^{tree}$ are shown.  \label{fig2}}
\end{figure}

\paragraph{Tree-level decay rate}
Two diagrams in Fig. \ref{fig1} contribute to the  tree level decay. 
If we artificially set the mass of the internal propagator $\mbox{M} \gg \mbox{m}$ (with electron mass \mbox{m}), 
the total $e^+e^-$-annihilation cross section and the p-Ps decay rate are calculated in the low 
momentum limit as
\begin{eqnarray}
\sigma_{0}(e^{+} e^{-} \rightarrow  \gamma \gamma)
&=& \frac{\pi r_{0}^{2}}{2 \beta}\frac{4r^{4}}{(1+r^{2})^{2}}, \label{tot2} \\
\Gamma(\mbox{p-Ps} \rightarrow \gamma \gamma) &=& (2 \beta) \cdot 4 \sigma_{0}(e^{+} e^{-} \rightarrow  \gamma \gamma) 
\cdot |\psi(0)|^{2}, \label{2decay}
\end{eqnarray}
where $\beta$ is the electron velocity, $r_{0} = \alpha_{e}/m $,
and $r=\mbox{m}/\mbox{M}$. Here the value of the positronium wave function at the origin is
$|\psi(0)|^{2}=1/\pi (2a_{0})^{3}$ with $a_{0}=1/(m\alpha_{e})$.
Making $r=1$, we obtain the well known p-Ps decay rate at the tree level. The same result can be
obtained using the optical theorem: 
\begin{eqnarray}
\sigma_{0}(e^{-}e^{+}  \rightarrow \gamma \gamma) = \frac{1}{4 \beta m^2} \mbox{Im}
 \bigl[\mbox{M}_{1}^{tree} + \mbox{M}_{2}^{tree} \bigr] , \label{2opti}
\end{eqnarray}
where two diagrams $\mbox{M}_{1}^{tree}$ and $\mbox{M}_{2}^{tree}$ are shown in Fig. \ref{fig2}.
However, we here expand the Wick rotated propagator by a large mass M before the loop integral as
$1/(l^{2}+\mbox{M}^{2})=\sum_{n=0}^{\infty}  (-l^{2})^{n} /\mbox{M}^{2(n+1)}$.
The right hand side of Eq. (\ref{2opti}) is calculated as 
\begin{eqnarray}
\mbox{Im} \bigl[\mbox{M}_{1}^{tree}  + \mbox{M}_{2}^{tree}  \bigr] = \pi \alpha_{e}^{2}
\bigl(8 r^4 -16 r^6 + \cdots \bigr),  \label{1tg}
\end{eqnarray}
which coincides with the first terms in the expansion of Eq. (\ref{tot2}). 
We introduce the gauge parameter in the photon propagator to check the gauge symmetry.
In general the introduction of the artificial mass M breaks the gauge symmetry, namely, the Ward-Takahashi identity.
However, when both of electron and positron  rest, the gauge symmetry is restored and the result in Eq. (\ref{1tg})
does not depend on gauge parameter.

\begin{figure}[tb]
\begin{center}
\includegraphics[width=14cm]{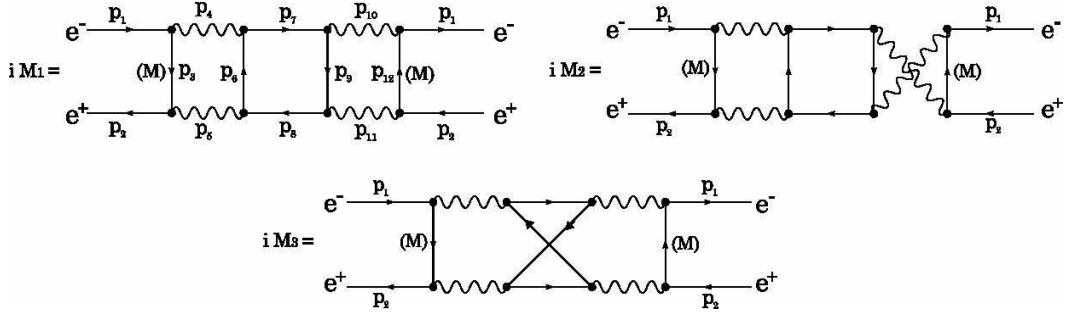}
\end{center} 
\caption{Three diagrams $\mbox{M}_{1}, \mbox{M}_{2}$ and $\mbox{M}_{3}$ are shown.\label{fig4} }
\end{figure}

\paragraph{Two-loop correction}
We proceed to two-loop contributions involving light-by-light  scattering diagrams.
We calculate the corrections $\Delta \sigma$ to the leading order cross section $\sigma_{0}$ through the optical theorem:
\begin{eqnarray}
\Delta \sigma  = \frac{1}{2 \beta m^{2}} \cdot \mbox{Im} \bigl[\mbox{M}_{1} +
 \mbox{M}_{2}+\mbox{M}_{3} \bigr],   \label{imm}
\end{eqnarray}
where the three-loop diagrams $\mbox{M}_{1}$, $\mbox{M}_{2}$, and $\mbox{M}_{3}$ are shown in Fig. \ref{fig4}.
Here the internal fermion loop is considered massless, while the case of electron in the internal loop 
has been considered in \cite{cza2}. The sum of imaginary parts of these diagrams is gauge-independent. 
For divergences, we employ the dimensional regularization with $D = 4 - 2\ep$.
The diagrams with inverse orientation of the fermion loop exist and their contributions are same. 
Assigning the artificial large mass to the two internal electron propagators, we now have to consider 
several possibilities for the loop momentum scales. 
In the case of diagram $\mbox{M}_{1}$, for example,  we take $p_{3}, p_{7},$ and $p_{12}$ as the three independent 
 loop momenta and the following four regions exist : 1.  $p_{3} \ll \mbox{M}, \ p_{7} \ll \mbox{M}, \ p_{12} \ll \mbox{M} $,
 2. $p_{3} \simeq \mbox{M}, \ p_{7} \ll \mbox{M}, \ p_{12} \ll \mbox{M} $, 3. $p_{3} \simeq \mbox{M},
 \ p_{7} \simeq \mbox{M},  \ p_{12} \ll \mbox{M} $, and 4. $p_{3} \simeq \mbox{M}, \ p_{7} \ll \mbox{M}, 
 \ p_{12} \simeq \mbox{M} $.
To calculate the hardest three-loop integrals in region 1 (all soft momenta), we use the MINCER package \cite{lar} 
and the factorized integrals are 
calculated in FORM \cite{ver} using well-known formulae.
Although each of $\mbox{Im} [\mbox{M}_{1}]$ and $\mbox{Im} [\mbox{M}_{2}]$ has the divergences of $1/\epsilon$
and depends on gauge parameter, the sum is free from divergences and gauge parameter. 
We can obtain it up to the first eight terms:
\begin{eqnarray}
\mbox{Im} \bigl[\mbox{M}_{1} + \mbox{M}_{2}\bigr] \cdot \frac{\pi}{\alpha^{4}} &=&
 + r^4  (
 -0.394 +4 \log r
          )   
       + r^6  (
          0.121 -8 \log r
        )  \nonumber  \\
   &+&   r^8  (
      0.113 +12.3 \log r
          )   
       + r^{10} (
     -0.316 -16.9 \log r
          )\nonumber  \\
   &+&  r^{12} (
     0.481 +21.7 \log r
          ) 
+ r^{14} (
     -0.618 -26.7 \log r
          ) \nonumber  \\
&+& r^{16} (
     0.734 + 31.9 \log r
          ) 
+ r^{18} (
     -0.834 - 37.3 \log r
          ). \label{2resm12}
\end{eqnarray}
Here and hereafter only three digits in the coefficients are shown.
$\mbox{Im} [\mbox{M}_{3} ]$ can be calculated up to the first six terms:
\begin{eqnarray}
\mbox{Im} \bigl[\mbox{M}_{3} \bigr] \cdot \frac{\pi}{\alpha^{4}} = 
 &+& r^4  (
       -0.313 +8 \log r
          )   
       + r^6  (
         -0.706  -16 \log r
          ) \nonumber  \\
       &+& r^8  (
       1.40  +25.0 \log r
          )  
       + r^{10}  (
        -1.93  -34.8  \log r
          ) \nonumber  \\
       &+& r^{12}  (
        2.31 + 45.2 \log r 
          ) 
+ r^{14}  (
        -2.62 - 56.1\log r 
          ). \label{2resm3} 
\end{eqnarray}
$\mbox{Im} \bigl[\mbox{M}_{3}\bigr]$ is gauge invariant in itself and has no divergence.
The series in Eqs. (\ref{2resm12}) and (\ref{2resm3}) apparently do not converge at $r = 1$. 
However, we can choose an expansion parameter $z$ instead of $r$ such that these series converge.
The comparison between Eqs. (\ref{tot2}) and  (\ref{1tg}) suggests an appropriate substitution, 
\begin{eqnarray}
z=\frac{2r^{2}}{1+r^{2}}.
\end{eqnarray}
The change of the variable reconstructs the series in Eq. (\ref{2resm12}) into the following series of z,
\begin{eqnarray}
&&\mbox{Im} \bigl[\mbox{M}_{1} + \mbox{M}_{2}\bigr] \cdot \frac{\pi}{\alpha^{4}} =
 + z^2  (
 -0.445 + 0.5 \log z
          )   
       + z^3  (
 0.166
        )  \nonumber  \\
       &&+ z^4  (
      0.0108 + 0.0111 \log z
          )   
       + z^{5} (
    -0.000899 + 0.00740  \log z
          )\nonumber  \\
	&&+ z^{6} (
  -0.00156 + 0.00462 \log z
          )  
+ z^{7} (
 -0.00110 + 0.00297 \log z
          ) \nonumber \\
&&+ z^{8} (
  -0.000690 + 0.00198 \log z
          )  
+ z^{9} (
 -0.000421 + 0.00137 \log z
          ),   \label{2zm12}  
\end{eqnarray}
which are good convergent series. The series in Eq. (\ref{2resm3}) are also converted to good convergent
series of $z$. Since the obtained series are limited in the first several terms, we
should estimate the errors with the remaining infinite series. We rewrite the series in Eq. (\ref{2zm12}) as 
$\sum_{i=2}^{\infty}c_{i}z^{i}$ and estimate the errors as 
$\biggl|\sum_{i=10}^{\infty}c_{i} \biggr| \ < \ |c_{9}| \sum_{i=1}^{\infty}t^{i}$.
Here we can conservatively set $t=0.62$ because of the ratio $c_{9}/c_{8}\simeq 0.610$. We can obtain the result,
\begin{eqnarray}
\mbox{Im} \bigl[\mbox{M}_{1} + \mbox{M}_{2}\bigr] \cdot \frac{\pi}{\alpha^{4}} &=&
\biggl(\sum_{i=2}^{9}c_{i} + \frac{tc_{9}}{2(1-t)} \biggr) \pm \frac{t|c_{9}|}{2(1-t)} \\
&=& -0.2726 \pm 0.00034 \ (\equiv d_{12}),  \label{2rese12}
\end{eqnarray}
where we write this result as $d_{12}$ for the latter references.
In the same way, we can obtain the results,
\begin{eqnarray}
\mbox{Im} \bigl[\mbox{M}_{3} \bigr] \cdot \frac{\pi}{\alpha^{4}} &=& -0.4692 \pm 0.0029 \  (\equiv d_{3}). \label{2rese3}
\\
\mbox{Im} \bigl[\mbox{M}_{1} + \mbox{M}_{2}+\mbox{M}_{3} \bigr]
\cdot \frac{\pi}{\alpha^{4}} &=& -0.7419 \pm 0.0033 \ (\equiv d_{tot}).  \label{2totres}
\end{eqnarray}
We check that the results in Eqs. (\ref{2rese12}) and (\ref{2rese3}) are supported
by the results of two summation methods, H\"{o}lder summation and Shanks transformation.

\section{Charmonium decay  \label{sec3}}
\begin{figure}[tb]
\begin{center}
\includegraphics[width=10cm]{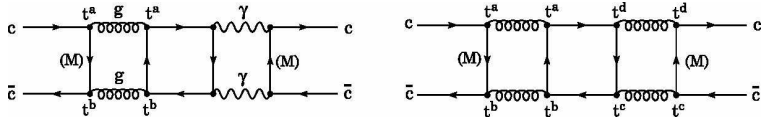}
\end{center} 
\caption{ Diagram corresponding to $\mbox{M}_{1}$ in Fig. \ref{fig4} for the decay
$\eta_{c} \rightarrow \gamma \gamma$ ($\eta_{c} \rightarrow gg$) is shown in the left (right).
\label{fig8} } 
\end{figure}

\paragraph{Paracharmonium decays to photons}
The decay rate is given in the analogous form to Eq. (\ref{2decay}) as
\begin{eqnarray}
\Gamma(\eta_{c} \rightarrow \gamma \gamma) = (2 \beta) \cdot 4 \sigma_{tot}(c \bar{c} \rightarrow
\gamma \gamma) \cdot |\psi_{c}(0)|^{2},  \label{3etad}
\end{eqnarray}
where $\psi_{c}$ is the charmonium wave function. The tree-level decay rate is
$\Gamma_{0}(\eta_{c} \rightarrow \gamma \gamma)=
4\pi  \alpha_{e}^{2} (Q_{c})^{4} N_{c} |\psi_{c}(0)|^{2}/m_{c}^{2}$
with the charm mass $m_{c}$, the electric charge of the charm quark $Q_{c}$, and
$N_{c}=3$. We here use the color singlet wave function for  $\eta_{c}$,
\begin{eqnarray}
| {\eta}_{c} \rangle = \frac{1}{\sqrt{N_{c}}} \ \delta_{ij} \  | c^{i} \bar{c}^{j} \rangle, \label{singlet}
\end{eqnarray}
where $i$ and $j$ are the color indices. Two-loop QCD corrections have the
three diagrams corresponding to $\mbox{M}_{1}$, $\mbox{M}_{2}$ and 
$\mbox{M}_{3}$ in Fig. \ref{fig4}. Diagram $\mbox{M}_{1}(c \bar{c} \rightarrow c \bar{c})$ 
corresponding to $\mbox{M}_{1}$ is shown in the left in Fig. \ref{fig8}, 
where $t^{a}$ is the generators of QCD.
We restrict the quarks running in the fermion loop to the up, down, and strange quarks
which can be regarded as massless. We can obtain $\mbox{Im} [\mbox{M}_{1}
(c \bar{c} \rightarrow c \bar{c})]$, counting the extra factors relative to 
$\mbox{Im} [\mbox{M}_{1}(e^{-}e^{+} \rightarrow e^{-}e^{+})]$, as
\begin{eqnarray}
\frac{\mbox{Im} \bigl[\mbox{M}_{1}(c \bar{c} \rightarrow c \bar{c} ) \bigr]}
{\mbox{Im} \bigl[\mbox{M}_{1}(e^{-}e^{+} \rightarrow e^{-}e^{+})\bigr]}
 =\biggl(\frac{\alpha_{s}}{\alpha_{e}}\biggr)^{2}
(Q_{c})^{2} [(Q_{u})^{2}+(Q_{d})^{2}+(Q_{s})^{2}] \cdot \mbox{T}_{R}^{2}(N_{c}^{2}-1),  \label{3ext}
\end{eqnarray}
where $\alpha_{s}$ is the strong coupling constant and $\mbox{T}_{R}=1/2$.
The extracted factors in Eq. (\ref{3ext}) are common to the diagram where the orientation of the fermion loop is reversed
in Fig. \ref{fig8} and they are also common to the other two diagrams corresponding to $\mbox{M}_{2}$ and 
$\mbox{M}_{3}$ in Fig. \ref{fig4}. Then, using Eqs. (\ref{imm}), (\ref{3etad}) and (\ref{3ext})
with $Q_{u}=Q_{c}=2/3$ and $Q_{d}=Q_{s}=-1/3$, we obtain the decay rate including the two-loop corrections as
\begin{eqnarray}
\Gamma(\eta_{c} \rightarrow \gamma \gamma) 
= \Gamma_{0}(\eta_{c} \rightarrow \gamma \gamma)
\biggl[1+\biggl(\frac{\alpha_{s}}{\pi}\biggr)^{2} d_{tot} \biggr].  \label{3degg}
\end{eqnarray}
\paragraph{Paracharmonium decays to gluons }
The decay rate is obtained by the formula analogous to Eq. (\ref{3etad}).
The tree-level decay rate is 
$\Gamma_{0}(\eta_{c} \rightarrow g g) =8\pi \alpha_{s}^{2} \mbox{T}_{R}^{2}C_{F}|\psi_{c}(0)|^{2}/m_{c}^{2}$
with $C_{F}=4/3$. Three diagrams which correspond to three in Fig. \ref{fig4} contribute to the two-loop QCD corrections. 
Diagram $\mbox{M}_{1}(c \bar{c} \rightarrow c \bar{c})$ is the right one in Fig. \ref{fig8}.
Using the wave function in Eq. (\ref{singlet}), we obtain the common extra factors for
 $\mbox{M}_{1,2}(e^{-}e^{+} \rightarrow e^{-}e^{+})$ as
\begin{eqnarray}
\mbox{Im} \bigl[\mbox{M}_{1,2}(c \bar{c} \rightarrow c \bar{c}) \bigr]  
= \mbox{Im} \bigl[\mbox{M}_{1,2}(e^{-}e^{+} \rightarrow e^{-}e^{+}) \bigr] \cdot \mbox{T}_{R}^{2} C_{F}^{2}  n_{q},  
\label{3ex12}
\end{eqnarray}
where $n_{q}$ is the number of the light quarks.
The color factor for $\mbox{M}_{3}$ is different from one for $\mbox{M}_{1}$ and $\mbox{M}_{2}$, 
as $\mbox{T}_{R}^{2} C_{F}\bigl(C_{F} - C_{A}/2 \bigr) n_{q}$
with $C_{A}=3$. Thus, we obtain the decay rate as 
\begin{eqnarray}
\Gamma(\eta_{c} \rightarrow g g)=\Gamma_{0}(\eta_{c} \rightarrow g g)
\biggl[1+\biggl(\frac{\alpha_{s}}{\pi}\biggr)^{2} \frac{n_{q}}{2} \biggl(C_{F} d_{tot} -\half C_{A} d_{3} \biggr) \biggr].
 \label{3dggres}
\end{eqnarray}
It should be noted that 
if we average the colors of the charm and anti-charm quarks in this calculation,
the divergences are not cancelled because the color factors for $\mbox{M}_{1}$ and  $\mbox{M}_{2}$ are 
different as $\mbox{Tr}[t^{a}t^{b}t^{c}t^{d}] \mbox{Tr}[t^{a}t^{d}t^{c}t^{b}] \not=
\mbox{Tr}[t^{a}t^{b}t^{c}t^{d}] \mbox{Tr}[t^{a}t^{d}t^{b}t^{c}]$.
The QCD bare lagrangian has the gluon four-point interaction, unlike QED, and the gluon four-point function has divergences
proportional to the interaction term in the bare lagrangian.
However, the requirement that $\eta_{c}$ is color singlet state in Eq. (\ref{singlet})
reduces the color factors for $\mbox{M}_{1}$ and  $\mbox{M}_{2}$ from the different ones to the common ones in Eq. (\ref{3ex12})
and eliminates the potential divergences in the two-loop QCD corrections.
\paragraph{ Numerical values of corrections}
We first estimate the corrections in $\eta_{c} \rightarrow \gamma \gamma$ as
\begin{eqnarray}
\Gamma(\eta_{c} \rightarrow \gamma \gamma) &=& \Gamma_{0}(\eta_{c} \rightarrow \gamma \gamma)
\biggl[1+\biggl(\frac{\alpha_{s}}{\pi}\biggr) \delta_{\gamma}^{(1)}+\biggl(\frac{\alpha_{s}}{\pi}\biggr)^{2}
\delta_{\gamma}^{(2)} \biggr]  \label{3epp}  \\
&=& \Gamma_{0}(\eta_{c} \rightarrow \gamma \gamma)
\bigl[1+(-0.439)+ (-0.0125 + 0.017 \cdot \delta_{\gamma}^{others}) \bigr]. \label{3epp2}
\end{eqnarray}
Here we use the values of the one-loop corrections $\delta_{\gamma}^{(1)} \simeq -3.38$ \cite{har, kwo}
and the two-loop corrections are written as $\delta_{\gamma}^{(2)} = d_{tot} + \delta_{\gamma}^{other}$
from Eq. (\ref{3degg}). $\delta_{\gamma}^{other}$ denotes the other two-loop corrections.
We also use the coupling constant at the scale of the charm quark mass, $\alpha_{s}(m_{c})=0.41$ \cite{bet}.
We find that the obtained two-loop corrections account for $- 1.25 \%$ of the tree-level decay rate,
neglecting the uncertainty of $d_{tot}$ in Eq. (\ref{2totres}) due to its smallness.
We next estimate the corrections in $\eta_{c} \rightarrow g g$ in the same way:
\begin{eqnarray}
\Gamma(\eta_{c} \rightarrow gg)  = \Gamma_{0}(\eta_{c} \rightarrow g g)
\bigl[1+(0.633)+ (-0.00729 + 0.017 \cdot \delta_{g}^{others}) \bigr]. \label{3gg}
\end{eqnarray}
where in the analogue to Eq. (\ref{3epp}), the one-loop correction is $\delta_{g}^{(1)} \simeq 4.87$ \cite{bar, hag, kwo}.
The two-loop correction is $\delta_{g}^{(2)} = n_{q}(C_{F} d_{tot} -C_{A} d_{3}/2)/2 + 
\delta_{g}^{other}$ from Eq. (\ref{3dggres}). 
The obtained corrections account for $- 0.73 \%$ of the tree level.
The ratio of the decay rates in Eq. (\ref{3epp2}) to one in Eq. (\ref{3gg}) is free from the ambiguity of the wave function 
$\psi_{c}(0)$. To compare the ratio to the experimental value, we need the other corrections 
$\delta_{\gamma}^{others}$ and $\delta_{g}^{others}$, which we leave for future work.

\section{Summary  \label{sec4}}
We calculate the two-loop QCD corrections to the decay modes $\eta_{c} \rightarrow
\gamma \gamma$ and $\eta_{c} \rightarrow g g$ involving light-by-light  scattering diagrams 
with light quark loops by the method of LME. 
The obtained series in LME are transformed into ones with good convergence properties
 by a change of the expansion parameter. 
 The obtained corrections to the decay $\eta_{c} \rightarrow \gamma \gamma$ account for $-1.25 \%$ of the tree-level
and the corrections to $\eta_{c} \rightarrow g g$ account for $ -0.73 \%$. 
The artificially introduced large mass does not break the gauge symmetry in the kinematic case
of both initial particles at rest because the results are independent of the gauge parameter.
Regarding the divergences of the two-loop QCD corrections to $\eta_{c} \rightarrow g g$ in the present paper, 
the nature that the observed hadron, $\eta_{c}$ in this case, is color singlet state, eliminates
the potential divergences.

\begin{acknowledgments}
We are grateful to A. Czarnecki for the suggestion of this subject and the helpful advices.
The work of K.H. is supported by the Science and Engineering Research, Canada.
\end{acknowledgments}

\end{document}